\documentclass{article}

\usepackage{arxiv}

\usepackage[utf8]{inputenc} 
\usepackage[T1]{fontenc}    
\usepackage{hyperref}       
\usepackage{url}            
\usepackage{booktabs}       
\usepackage{amsfonts}       
\usepackage{nicefrac}       
\usepackage{microtype}      
\usepackage{lipsum}
\usepackage{graphicx}
\usepackage{subcaption}
\usepackage{amssymb, amsmath}
\usepackage[numbers]{natbib}
\graphicspath{ {./images/} }

\title{The impact of abnormal temperatures on crop yields in Italy: a functional quantile regression approach.}

\author{
 Giovanni Bocchi \\
  Department of Environmental Science and Policy\\
  Università degli Studi di Milano\\
  Via Celoria 10, Milano, 20133 \\
  \texttt{giovanni.bocchi1@unimi.it} \\
   \And
 Alessandra Micheletti \\
  Department of Environmental Science and Policy\\
  Università degli Studi di Milano\\
  Via Celoria 10, Milano, 20133 \\
  \texttt{alessandra.micheletti@unimi.it} \\
  \And
 Paolo Nota \\
  Department of Environmental Science and Policy\\
  Università degli Studi di Milano\\
  Via Celoria 10, Milano, 20133 \\
  \texttt{paolo.nota@unimi.it} \\
  \And
Alessandro Olper \\
  Department of Environmental Science and Policy\\
  Università degli Studi di Milano\\
  Via Celoria 10, Milano, 20133 \\
  \texttt{alessandro.olper@unimi.it} \\
}

\date{}

\begin{document}
\maketitle
\begin{abstract}
In this study, we apply functional regression analysis to identify the specific within-season periods during which temperature and precipitation anomalies most affect crop yields. Using provincial data for Italy from 1952 to 2023, we analyze two major cereals, maize and soft wheat, and quantify how abnormal weather conditions influence yields across the growing cycle. Unlike traditional statistical yield models, which assume additive temperature effects over the season, our approach is capable of capturing the timing and functional shape of weather impacts. In particular, the results show that above-average temperatures reduce maize yields primarily between June and August, while exerting a mild positive effect in April and October. For soft wheat, unusually high temperatures negatively affect yields from late March to early April. Precipitation also exerts season-dependent effects, improving wheat yields early in the season but reducing them later on. These findings highlight the importance of accounting for intra-seasonal weather patterns to provide insights for climate change adaptation strategies, including the timely adjustment of key crop management inputs.
\end{abstract}

\keywords{Functional regression, fPCA, Quantile regression, climate change, Italy}

\section{Introduction}\label{sec1}

Statistical tools have long been used to estimate the relationship between weather conditions and agricultural outcomes \citep{Hodges1931}, and they are now widely applied in the field \citep{Ortiz-Bobea2021}. A more recent and increasingly important application of these methods is the estimation of climate change impacts on crop productivity \citep{Hultgren2025}.

Most of the traditional statistical yield models, however, overlook the timing of weather conditions and typically assume that temperature effects on cereal yields are additive over the growing season. Some studies attempt to distinguish seasonal patterns, but these efforts remain limited in scope and flexibility \citep{Ortiz-Bobea2019, Shew2020}.

At the same time, Functional Data Analysis (FDA)~\citep{Ramsay2005} has been developed specifically for the statistical analysis of data characterized by complex functional dependencies, such as temporal and seasonal dependencies. In the broad field of Functional Data Analysis, functional regression models allow us to interpret the impact of some functional covariates, such as historical series of climatic variables, on other variables, possibly scalar, such as the annual production of crops. Indeed, a functional regression model allows estimation of coefficients that, having the same functional dependence as the data, permit a precise interpretation of the impacts throughout the period of interest.

For this reason, in this study, we employ a functional linear regression analysis to investigate specific within-season periods during which temperatures, with a specific focus on abnormal temperatures due to climate change, affect crop yields. Using provincial-level data for Italy from 1952 to 2023 as an extensive case study, we are able to clearly detect the parts of the growing season that are most sensitive to temperature and precipitation anomalies. We focus on two major crops: maize and soft wheat. Our findings indicate that above-average temperatures contribute to reducing maize yields during the June–August period, while exerting a positive effect during the early and late parts of the season (April and October). For wheat, unusually high temperatures have a negative impact on yields, particularly from late March to early April. Regarding precipitation, higher levels improve wheat yields early in the season but have detrimental effects in the later stages.

These results may have implications for climate change adaptation and may support more precise adjustments of key inputs, such as irrigation, at the moments when they are most needed.

The study is organized as follows. Section \ref{sec2} discusses the relevant literature regarding crop yield models and Functional Data Analysis in greater detail; Section \ref{sec3} outlines the functional regression framework we employed; Section \ref{sec4} presents the results of the application to Italian maize and wheat production data; and Section \ref{sec5} concludes.

\section{Background}\label{sec2}
\subsection{Background on yield models}

By now, it is well recognized that most statistical yield models overlook the timing of weather conditions and typically assume that temperature effects on cereal yields are additive during the growing season. In these frameworks, the specific moment when abnormal temperatures occur is treated as irrelevant. For example, the seminal study by \citep{Schlenker2009} assumes additive separability of temperature effects and defines a fixed growing season (e.g., March–August for corn), regressing aggregate weather in those months on annual yields.

Agronomic research, however, emphasizes that the timing of weather shocks is crucial for plant development \citep{Fageria2006}. Critical phenological phases, such as flowering, strongly influence the ability of a plant to accumulate biomass, making weather shocks during these periods particularly damaging.

One simple way to relax the additivity assumption is to include monthly or seasonal weather variables within the growing season. The same paper mentioned above by \citep{Schlenker2009} shows that July is the month where high temperatures affect the most U.S. corn yields (although model fit does not improve), coinciding with the flowering phase (see their Appendix). \citep{Gammans2017} shows that French wheat and barley respond differently across seasons, with warmer temperatures reducing yields during the warm months but having no or a positive effect in the fall. In the Italian context, \citep{Bozzola2018}, using a Ricardian cross-sectional approach, find that abnormal summer temperatures depress agricultural land values, while winter warmth has a positive effect. In the same context, using a panel data approach, \citep{Baldoni2025} shows heterogeneous weather effects across seasons on farms' productivity. However, a limitation of this strategy is its reliance on calendar months rather than agronomically defined crop stages.

Other studies explicitly incorporate the biophysical structure of crop development. \citep{Ortiz-Bobea2013} examines temperature effects across three phenological sub-seasons of U.S. corn and shows that plausible adaptation measures, such as shifting planting dates, could substantially mitigate projected yield losses from higher temperatures. Similarly, \citep{Shew2020} links weather conditions with wheat development stages in South Africa to assess cultivar-specific sensitivity to extreme heat, suggesting that sharing gene pools across breeding programs may help reduce vulnerability.

Other recent works by \citep{Ortiz-Bobea2019} and \citep{Ortiz-Bobea2021} allow the marginal effects of temperature to vary smoothly throughout the growing season. Using a two-dimensional spline, they show that for U.S. corn, higher temperatures early in the season (e.g. April) can increase yields, while heat during June–August is strongly detrimental. For winter wheat, \citep{Ortiz-Bobea2019} find that the most damaging temperature increases occur in spring, particularly in March. 

Only a limited number of economic studies have examined how weather variability affects Italy’s agricultural sector. \citep{Chavas2019} analyzes wheat and maize yields for seven provinces from 1900 to 2014 and shows that rising temperatures exert disproportionately strong negative effects on the lower tail of the yield distribution. Their findings suggest that low-yield regions or periods, often associated with less advanced technologies, are particularly vulnerable to heat stress. Building on this work, \citep{Chavas2022} investigates the spatial and temporal evolution of yield and production risks for corn and wheat using a similar dataset, linking these patterns to implications for national food security.

Other contributions focus specifically on durum wheat. \citep{Tappi2022} find that durum yields decline with higher precipitation and follow an inverted-U relationship with maximum temperatures. In a subsequent paper, \citep{Tappi2023} explores how weather shocks at different stages of the growing season affect yields across durum wheat varieties.

A broader, sector-level perspective is provided by \citep{Bozzola2018}, who employ a Ricardian (cross-sectional) framework with farm-level data and show that higher temperatures reduce agricultural land values, especially for crop-intensive and irrigated farms such as maize producers. \citep{Baldoni2025} assesses the impact of weather anomalies on the total factor productivity (TFP) of Italian farms across calendar seasons. They find that field crop farmers are particularly sensitive to rainfall during spring and summer, but not to temperature anomalies.
Finally, \citep{Olper2021}, using provincial panel data for 1980–2014, find that temperature shocks depress agricultural gross value added per worker, with poorer provinces suffering more severe impacts.

Despite these contributions, the existing literature pays little attention to the timing of weather conditions, an issue that lies at the core of our analysis.

\subsection{Background on Functional Data Analysis}

In this perspective, our analysis is grounded in the framework of Functional Data Analysis (FDA)~\citep{Ramsay2005, ferraty2006nonparametric, horvath2012inference, kokoszka2017introduction, cuevas2014overview}. Functional Data Analysis has emerged as a powerful methodology for investigating complex functional dependencies within data, thereby extending beyond the many limitations of multivariate approaches. In the FDA context, discrete data points measured repeatedly over a particular domain, whether temporal or spatial in nature, are usually smoothed using a basis expansion procedure to effectively obtain their functional representations.

Once a dataset consisting of discretized observations of functional objects defined on continuous domains has been obtained, the FDA toolbox enables a range of analyses that serve as generalizations of their multivariate counterparts. For instance, Functional Principal Component Analysis (fPCA)~\citep{shang2014} can be employed to identify the most important modes of variation present in the data. The difference is that the estimated Principal Components are functional objects as well, and may be very likely interpretable within the domain of interest. Analogously to its multivariate analogue, fPCA can be utilized to reduce the dimensionality of the data by projecting it onto the functional Principal Components. The resulting scores, obtained through this process, can then be leveraged in a great variety of applications for clustering or classification purposes. For irregularly sampled or sparse trajectories, PCA methods based on conditional expectations (PACE) are widely used \citep{yao2005functional}, and principal component modeling for sparse functional data dates back at least to \citep{james2000pca}.

Another significant class of tools in the context of FDA is functional regression models~\citep{reiss2017, morris2015functional, TALSKA2018}, which investigate the linear dependence between a functional or scalar target variable and one or more explanatory variables, where these variables can comprise both scalars and functions. A key advantage of functional regression models is that when the response is scalar and a functional covariate is present, the regression coefficient itself becomes a functional object defined on the same domain of interest. This allows for the evaluation of the effect of the covariate across its entire domain of interest. For instance, if the domain of interest corresponds to a temporal interval, such as the growing season of a particular crop, the model enables the estimation of the covariate's time-varying effect, thus accounting for the timing of the covariate values.

Furthermore, when the response variable has a scalar nature, recent developments have integrated the functional linear model with the quantile regression approach~\citep{koenker1978, Koenker2005}, enabling estimates of the regression coefficient that account for the domain-varying effect of the covariate on the distributional tails rather than just the mean. This technique, known as functional quantile regression~\citep{kato2012, li2022}, is particularly valuable in situations where there is interest in investigating the impact of functional covariates on specific regions of the scalar response variable's distribution, such as the upper or lower tails.
For instance, when analyzing yield data, functional quantile regression can be used to investigate the effect of functional covariates on the extreme values of the distribution.

Given the importance of understanding the temporal relationships between weather variables and crop yields, the methods outlined above are utilized to examine the specific impact of temperature and other weather conditions on agricultural productivity in Italy.

\section{Functional Regression model}\label{sec3}

\subsection{The scalar-on-function model}

To investigate the influence of extreme temperatures during the growing season on crop yields across various regions in Italy, we developed a functional scalar-on-function regression model. Such a model requires capturing the effects of at least three distinct factors: a global inter-seasonal trend that influences yields over longer time scales, spatial heterogeneity measured at the provincial level (Europe NUTS 3 classification), and, most pertinent to our study, the intra-seasonal impact of some weather covariates, specifically temperature and precipitation patterns.

Prior to specifying the model and outlining the estimation procedure, we fix the notation convention used throughout this study.

\begin{itemize}
	\item $i = i_0, \dots, i_0+N$ is the year index.
	\item $j = 1, \dots, M$ is the province index.
	\item $t \in [0, T]$ represents the crop's growing season, which remains consistent across all years but may differ among various crop types.
	\item $\verb|Yield|_{i,j}$ represents the crop yield in province $j$ during year $i$, and is calculated as the quotient of total production and harvested area.
	\item $Y_{i,j} = \log(\verb|Yield|_{i,j})$ is the natural logarithm of $\verb|Yield|_{i,j}$ which is taken as response variable.
	\item $\verb|X|^k_{i,j}(t)$ is the $k$-th weather covariate measured in province $j$ during year $i$ and evaluated at time $t$ of the growing season.
	\item $\verb|X|^k_{\bullet,\bullet}(t)$ is the average of the $k$-th weather covariate $\verb|X|^k$ over provinces and years, evaluated at time $t$ of the growing season.
\end{itemize}

With the established notation conventions, we now proceed to present the functional regression model used to analyze the crop production data:

\begin{equation}
	\label{eqn:model}
	Y_{i,j} = \alpha_j + \sum_{l=1}^B \beta_l (i - i_0)^l + \sum_{k=1}^G \int_{0}^T \gamma^k(t) (\verb|X|^k_{i,j} - \verb|X|^k_{\bullet,\bullet})(t) dt + \varepsilon_{i,j}
\end{equation}

In the above equation, the parameters to be estimated are the scalar quantities $\alpha_j$ and $\beta_l$, along with the functional parameter $\gamma^k(t)$, whereas $\varepsilon_{i,j}$ represents a zero-mean error term.

Examining Equation~\ref{eqn:model}, we observe that it can be partitioned into two distinct components, each serving a different purpose. The first component, represented by the non-functional terms $\alpha_j + \sum_{l=1}^L \beta_l (i - i_0)^l$, combines two key features: spatial heterogeneity among provinces through the use of fixed effects, $\alpha_j$, one for each province, and a slower temporal trend assumed to be uniform across provinces and polynomial in nature with respect to the year index $i$. In contrast, the second component, which is functional in nature, is where we determine the contribution of the $k$-th weather covariate through two successive steps: centering, followed by integration against a functional coefficient $\gamma^k(t)$. This approach is standard due to the scalar nature of the response variable $Y_{i,j}$.

\subsection{Model reduction}

To estimate the parameters of the model specified in Equation~\ref{eqn:model}, we employ a well-established technique based on functional Principal Component Analysis (fPCA), also referred to as the Karhunen-Loève transform or data-driven basis. Specifically, each centered climate covariate $\verb|X|^k$ is expanded using the basis provided by the eigenfunctions $\{\phi_s^k(t)\}_{s \ge 1}$ derived from the empirical covariance operator.

\begin{equation}
	\label{eqn:fpca}
	\verb|X|^k_{i,j}(t) - \verb|X|^k_{\bullet,\bullet}(t) = \sum_{s=1}^{\infty} \xi^{i,j,k}_s \phi_s^k(t)
\end{equation}

Here, the coefficients $\xi^{i,j,k}_s$ are the fPCA scores.

We then substitute the decomposition specified by Equation~\ref{eqn:fpca} into the model represented by Equation~\ref{eqn:model}, yielding the following expression:

\begin{equation}
	\label{eqn:model2}
	Y_{i,j} = \alpha_j + \sum_{l=1}^B \beta_l (i - i_0)^l + \sum_{k=1}^G \sum_{s=1}^{\infty} \xi^{i,j,k}_s \gamma^k_s + \varepsilon_{i,j}
\end{equation}

We denote the $s$-th coefficient of the basis expansion of the functional parameter $\gamma^k(t)$ on the empirical fPCA basis $\{\phi_s^k(t)\}_{s \ge 1}$ ad $\gamma^k_s = \int_{0}^T \gamma^k(t)\phi_s^k(t) dt$. Following this transformation, the functional dependence on $t$ has been eliminated in Equation~\ref{eqn:model2}. 

In principle, except for the infinite number of covariates, the model could now be estimated as a multiple regression. However, to proceed with estimation, we employ the standard choice of truncating each series to the first $L_k$ fPCs that collectively explain at least a proportion $\delta \in (0,1)$ of the total variance of the data. Having made this choice, the model specified in Equation \ref{eqn:model3} can be effectively estimated as a standard multiple regression using the fPCA scores $ \xi^{i,j,k}_s$ as additional scalar covariates.

\begin{equation}
	\label{eqn:model3}
	Y_{i,j} = \alpha_j + \sum_{l=1}^B \beta_l (i - i_0)^l + \sum_{k=1}^G \sum_{s=1}^{L_k} \xi^{i,j,k}_s \gamma^k_s + \varepsilon_{i,j}
\end{equation}

\subsection{Estimation}

The $M +B + \sum_{k=1}^G L_k$ parameters of the model specified by Equation~\ref{eqn:model3} can be estimated using Ordinary Least Squares (OLS), when the goal is obtaining estimates for the conditional expected value of the response variable $\mathbb{E}[Y | \verb|X|^1,\dots,\verb|X|^G]$. Additionally, the model parameters can also be estimated using a procedure rooted in the quantile regression (QR) framework. In this case, we are interested in estimating not only the conditional expected value but also the conditional quantile function $Q(Y, \tau | \verb|X|^1,\dots,\verb|X|^G)$ of the response variable for some given values of the quantile $\tau \in (0,1)$. By doing this, we can study not only the average behavior of the conditional target distribution but also its quantiles, particularly those related to the tails of the distribution. This aspect is of interest from an economic and policy perspective, given the heterogeneity of Italian crop producers. 

Based on this discussion, we consider the following estimates:

\begin{itemize}
	\item The OLS estimates of the coefficients: $\hat{\alpha}_j^{OLS}, \hat{\beta}_l^{OLS}$ and $\hat{\gamma}_s^{k, OLS}$.
	\item The simple QR estimates of the coefficients, fixed a value for the quantile $\tau \in (0,1)$: $\hat{\alpha}_j^{QR}(\tau), \hat{\beta}_l^{QR}(\tau)$ and $\hat{\gamma}_s^{k, QR}(\tau)$
	\item The Quantile Average (QA) estimates~\citep{koenker1984, wang2016} of the coefficients given a value of the quantile $\tau \in (0,1)$: $\hat{\alpha}_j^{QA}(\tau), \hat{\beta}_l^{QA}(\tau)$ and $\hat{\gamma}_s^{k, QA}(\tau)$
\end{itemize}

The QA estimates can be obtained by computing the average of simple QR estimates for quantile values located within a neighborhood $\mathcal{N}(\tau)$ centered around the quantile of interest $\tau$:

\begin{equation}
	\hat{\gamma}_s^{k, QA}(\tau) = \frac{1}{|\mathcal{N}(\tau)|}\sum_{\rho \in \mathcal{N}(\tau)} \hat{\gamma}_s^{k, QR}(\rho)
\end{equation}

This estimator has been proposed with the goal of reducing the variance of the simple QR estimator. However, the feasibility of this procedure relies on the hypothesis that the simple QR estimates exhibit minimal variation within the neighborhood of interest. This assumption is not universally valid and, indeed, ~\citep{li2022} have developed an adjusted Wald test to assess the equality of coefficients across quantile values in the neighborhood.

In conclusion, for every estimation procedure $e \in \{ OLS, QR, QA\}$, it is possible to reconstruct the functional profile of the estimated functional coefficients $\hat{\gamma}^{k, e}(t)$ associated with each weather covariate by leveraging the basis expansion and plug-in:

\begin{equation}
	\hat{\gamma}^{k, e}(t, \tau) = \sum_{s = 1}^{L_k}  \hat{\gamma}^{k, e}_s(\tau)\phi_s^k(t)
\end{equation}

To assess the uncertainty associated with our estimates, we utilize a Bootstrap resampling procedure. In each scenario, we generate $n_b$ Bootstrap replicas by re-sampling over provinces while maintaining the full temporal dynamics of the years, thereby enabling us to compute confidence bands for the functional parameters $\hat{\gamma}^{k, e}(t, \tau)$. 

The resulting estimates of these functional coefficients, along with their corresponding uncertainties, are subject to further investigation in the subsequent Section. Specifically, we quantify the impact of weather variables, particularly temperature, on crop production within the Italian agricultural system.

\section{Application to Italian crop yields data}\label{sec4}

In this Section, we apply the model and estimation procedures introduced previously to Italian crop production data for two specific crops: maize and wheat (common/soft). First of all, we introduce the dataset used for estimation. Subsequently, we present the results of the regression analysis, focusing on the functional coefficients associated with weather covariates, with particular attention devoted to the temperature coefficient. 

\subsection{Dataset}

We compile a new dataset that tracks annual production and cultivated area for maize and wheat across Italian provinces from 1952 to 2023. The data is sourced from official statistics provided by ISTAT (Italian National Institute of Statistics), with recent figures (2006-2023) obtained directly from the ISTAT website, and earlier data drawn from archival reports.

Our focus on maize and wheat is driven by their significant role in Italy's agricultural landscape, both in terms of production volume and cultivated area. The dataset spans 92 provinces, defined according to their 1952 boundaries to ensure consistent temporal coverage, despite changes in the number of provinces during the study period. This approach guarantees consistency in the data across time. Only provinces with complete data from 1952 onward are included, resulting in two balanced panels for 72 years: 79 provinces for maize and 68 for wheat. Yield is calculated as the ratio of production to cultivated area.

To explore the relationship between crop yields and weather conditions, we match our dataset with hourly weather data from ERA5 Land, covering the period from 1952 to 2023. ERA5 Land is a global gridded dataset with a resolution of 0.1$^\circ \times$ 0.1$^\circ$ ($\sim$9 km)~\citep{MunozSabater2019}. We focus on temperature and precipitation variables. To align the weather data with crop areas, we use a high-resolution crop type map for Europe from 2018 ($\sim$10 m resolution) to identify maize and wheat cultivation zones~\citep{dAndrimont2021}. We then select representative points within each province, gather the weather data for these locations, and compute the provincial averages. Further adjustments are made using high-resolution climatology data ($\sim$900 m)~\citep{Crespi2018, Brunetti2014} to ensure accurate representation of weather conditions in areas where crops are grown.

We consider hourly temperature and daily total precipitation over the crop-specific season. We defined the growing season for maize as spanning from April to October. For wheat, we consider the latest part of the growing season, occurring between February and June, as this period is associated with the greatest impact of heat \citep{Gammans2017}. Consequently, we have access to $N_T = 5136$ temperature observations during the maize growing season and $N_T = 3600$ temperature observations during the wheat growing season. For precipitation data, which are employed exclusively in the wheat analysis (as maize is fully irrigated), we have a dataset comprising $N_T = 150$ daily observations

We present now a series of plots to illustrate the meaning of the model terms introduced in Equation~\ref{eqn:model} and demonstrate their necessity. We begin by examining the temporal trends in crop yields across provinces, plotting the evolution of yields over years for each province.

\begin{figure}
	\centering
	\begin{subfigure}{0.49\textwidth}
		\includegraphics[width=\textwidth]{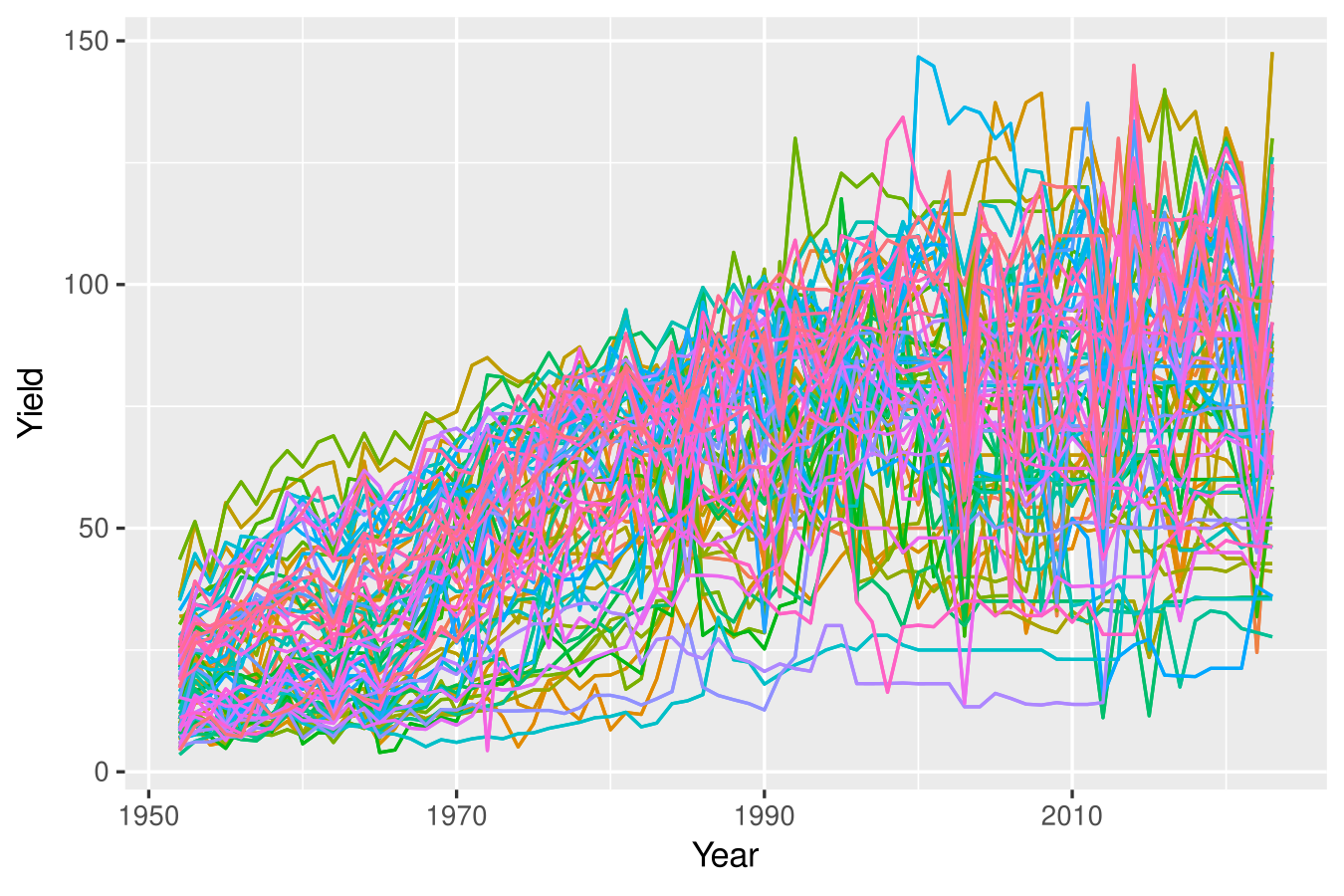}
		\caption{}
		\label{fig:yieldsmaize}
	\end{subfigure}
	\hfill
	\begin{subfigure}{0.49\textwidth}
		\includegraphics[width=\textwidth]{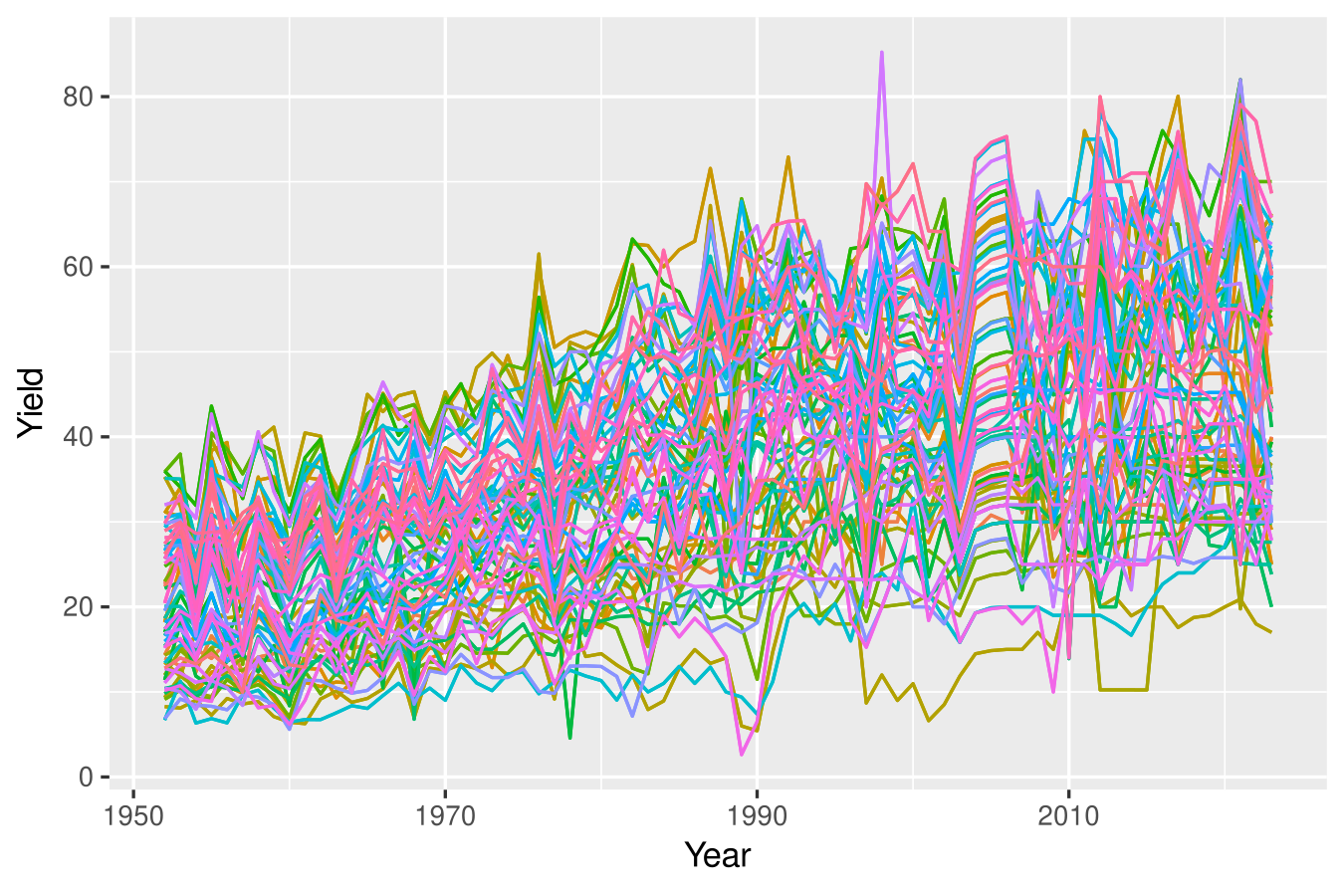}
		\caption{}
		\label{fig:yieldswheat}
	\end{subfigure}
	\caption{Figures illustrating the yield series (kg/ha) for each province, which are presented over the period from 1952 to 2023. Specifically, panel (a) depicts the temporal evolution of yields for maize, while panel (b) shows the corresponding evolution for wheat.}
	\label{fig:yieldsevolution}
\end{figure}

The two panels of Figure~\ref{fig:yieldsevolution} clearly illustrate that both crops exhibit a global positive trend over the bigger time scale of years, accompanied by a leveling off in yields after approximately the year 2000. This observed pattern motivated our decision to incorporate long-term year dependence into the model through the use of a polynomial function of degree $B=2$ in the year index variable $i$.

\begin{figure}
	\centering
	\begin{subfigure}{0.49\textwidth}
		\includegraphics[width=\textwidth]{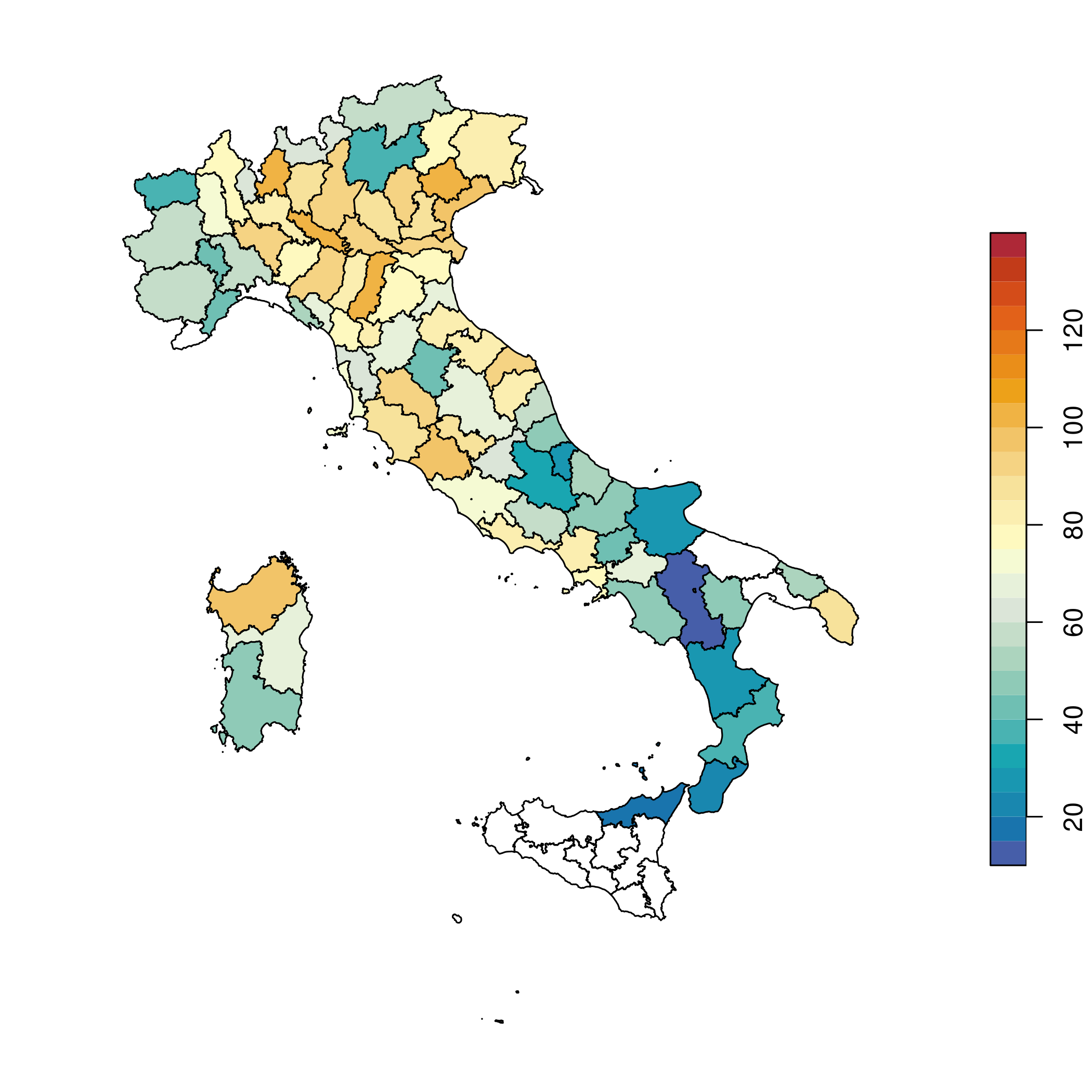}
		\caption{}
		\label{fig:spatialmaize1}
	\end{subfigure}
	\hfill
	\begin{subfigure}{0.49\textwidth}
		\includegraphics[width=\textwidth]{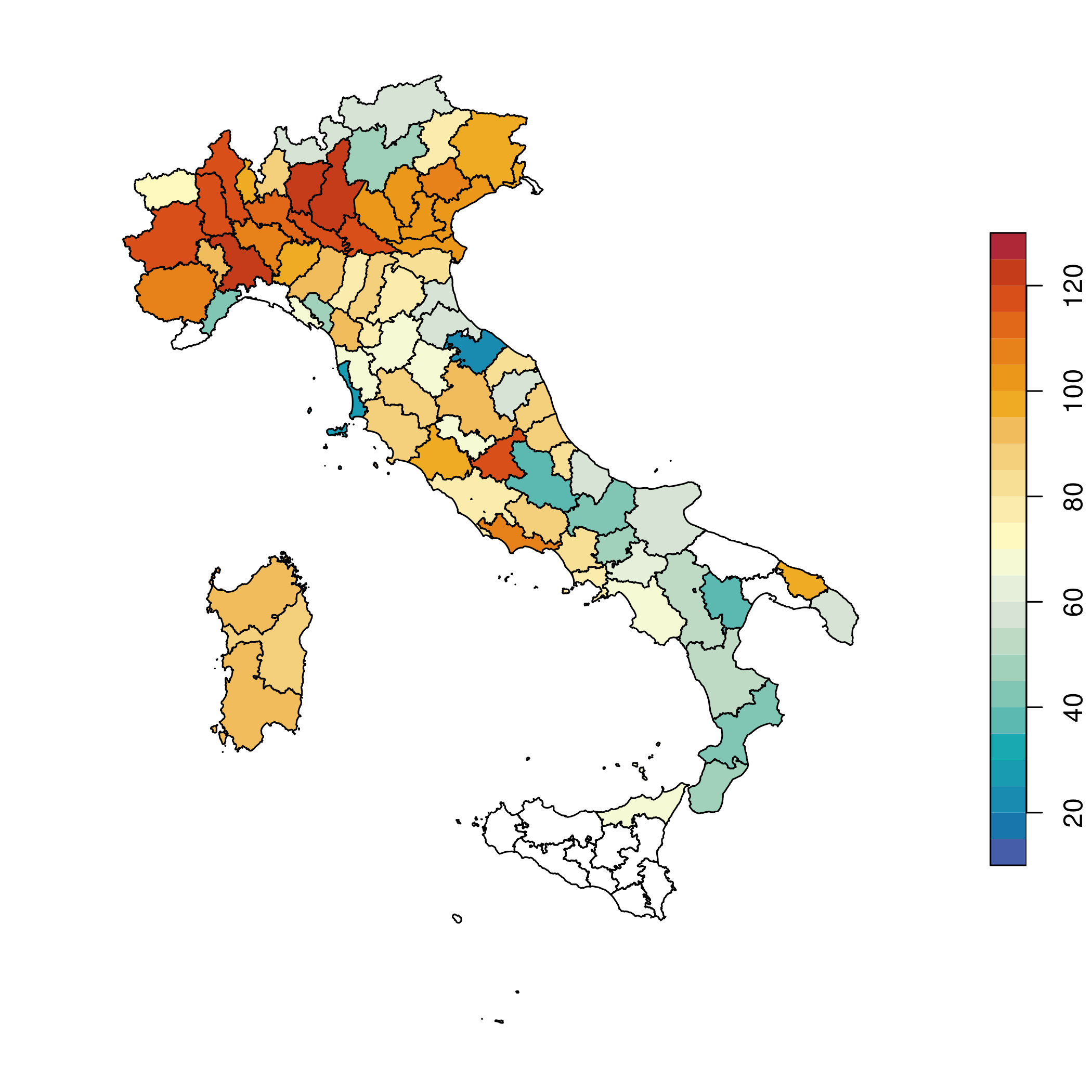}
		\caption{}
		\label{fig:spatialmaize2}
	\end{subfigure}
	\caption{The figure presents the spatial distribution of yield values (kg/ha) across provinces, highlighting regional variations in productivity. Specifically, panel (a) displays maize yields for the year 1990, while panel (b) shows the corresponding maize yields for 2021. Provinces depicted in white on each panel do not have available production data for that particular year.}
	\label{fig:spatialmaize}
\end{figure}

In addition to examining temporal trends, we can also visualize spatial patterns of yield variation across Italian provinces at different points in time. Figure~\ref{fig:spatialmaize} illustrates the spatial distribution of maize yields for two distinct years: 1990 and 2021. Notably, both maps reveal a concentration of maize production in northern Italy, particularly in the Po Valley, and parts of central Italy. A comparison of these maps, which share a common scale, further underscores the existence of a global trend towards increased yields, as evidenced by the prevalence of red tones on the 2021 map. 
Notably, such yield increases occurred despite the introduction in 2015 of the so-called \emph{greening} measures in the European Common Agricultural Policy, which resulted in a significant decrease in the land cultivated with maize as the main crop in Italy, particularly in the Lombardy region~\citep{Bertoni2018,Bertoni2021,Micheletti2020}.

\begin{figure}[htp]
	\centering
	\begin{subfigure}{0.49\textwidth}
		\includegraphics[width=\textwidth]{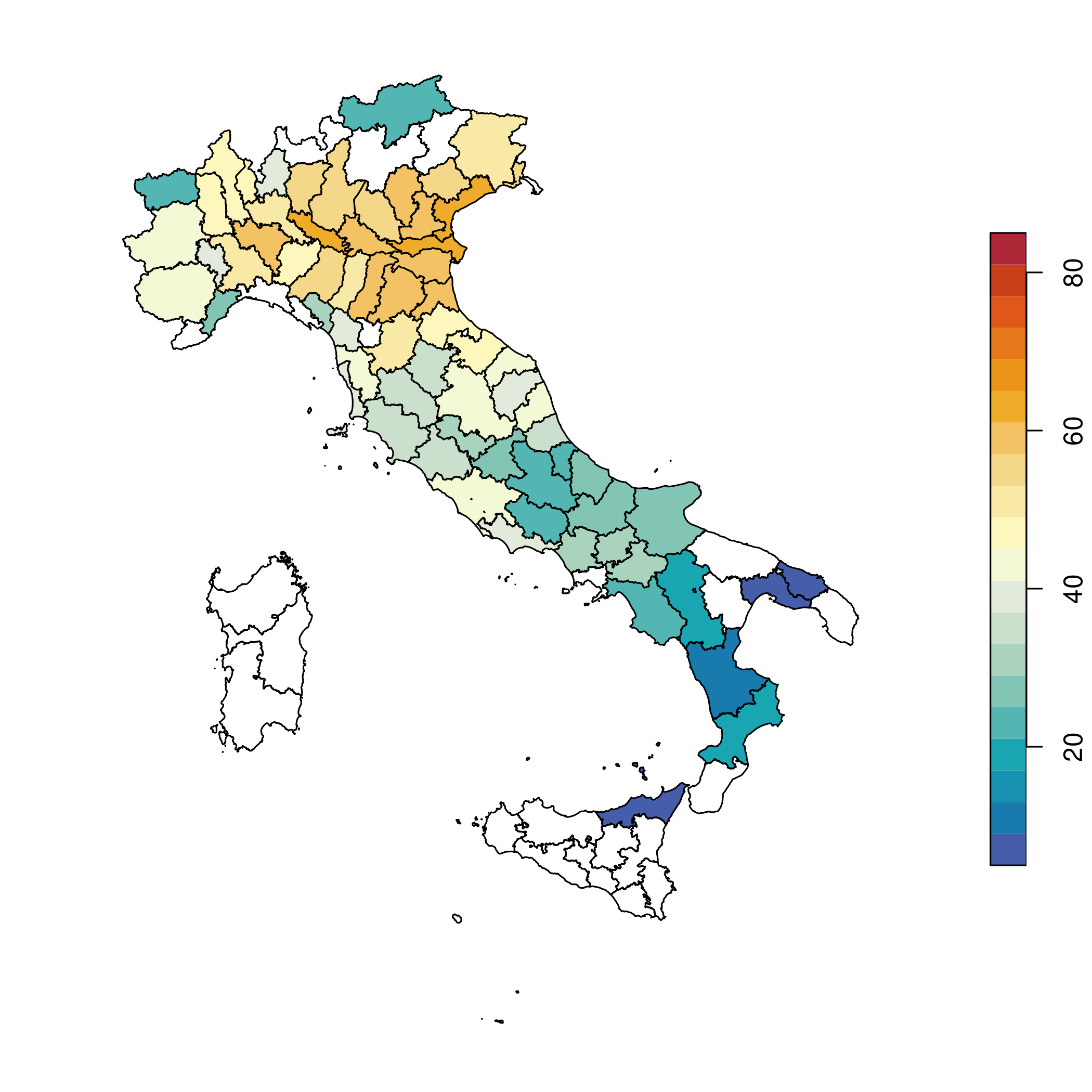}
		\caption{}
		\label{fig:spatialwheat1}
	\end{subfigure}
	\hfill
	\begin{subfigure}{0.49\textwidth}
		\includegraphics[width=\textwidth]{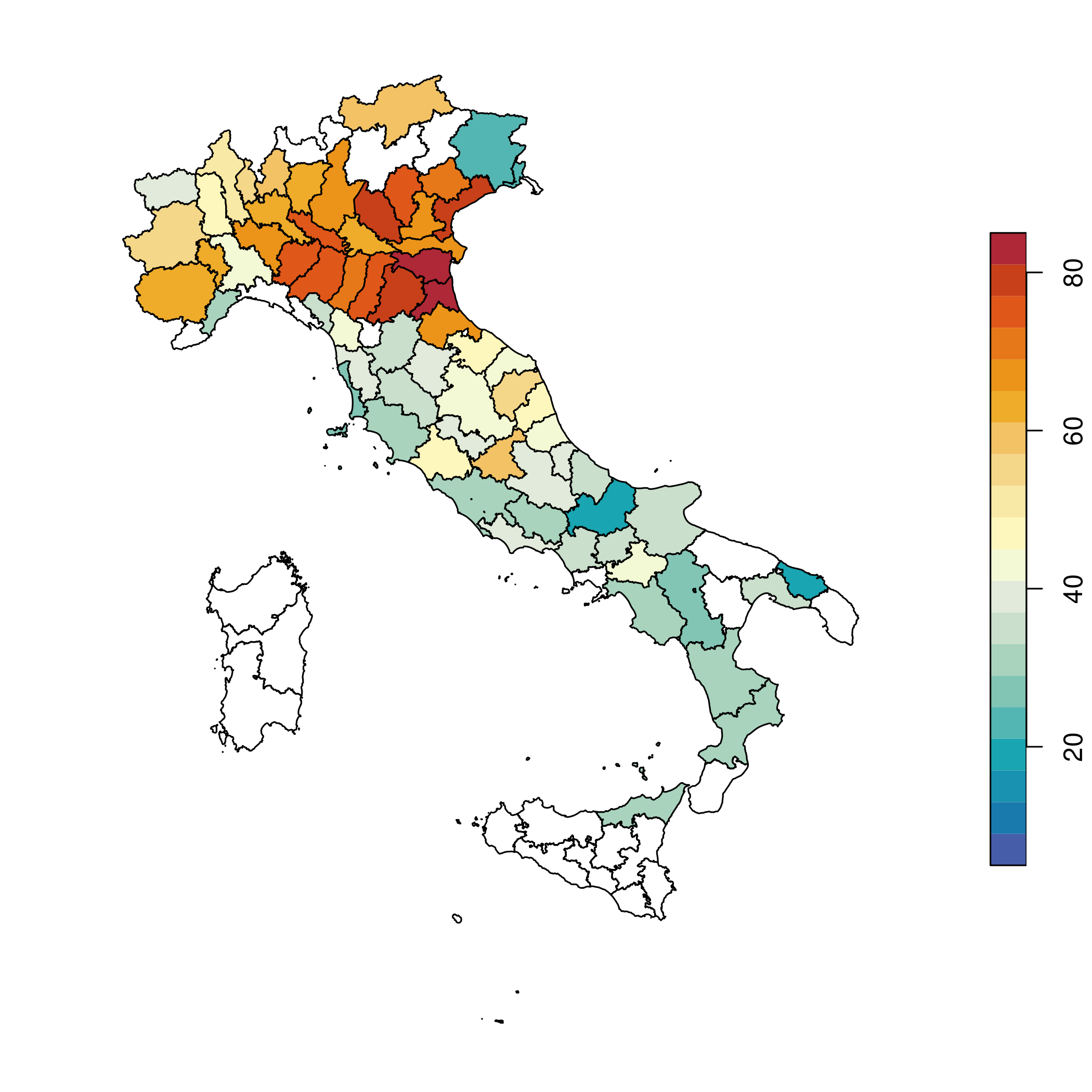}
		\caption{}
		\label{fig:spatialwheat2}
	\end{subfigure}
	\caption{The figure presents the spatial distribution of yield values (kg/ha) across provinces, highlighting regional variations in productivity. Specifically, panel (a) displays wheat yields for the year 1990, while panel (b) shows the corresponding wheat yields for 2021. Provinces depicted in white on each panel do not have available production data for that particular year.}
	\label{fig:spatialwheat}
\end{figure}

The Figure~\ref{fig:spatialwheat} allows for drawing similar observations regarding the distribution of wheat production. 

\subsection{Maize yields analysis}

We commence our analysis with an examination of maize yields data, taking into account the fact that maize is typically an irrigated crop. In order to isolate the effects of high temperatures on yield, we deliberately excluded precipitation from our analysis. As previously noted, a polynomial function of degree $B=2$ in the year variable $i$ was specified for both maize and wheat yields. Consequently, the model fitted to the maize yields data takes the form
\begin{equation}
	\label{eqn:modelmaize}
	Y_{i,j} = \alpha_j + \beta_1 (i - i_0)+ \beta_2 (i - i_0)^2 + \int_{0}^T \gamma(t) (\verb|Temp|_{i,j} - \verb|Temp|_{\bullet,\bullet})(t) dt + \varepsilon_{i,j}
\end{equation}

We expanded the temperature data on a Fourier basis due to its inherent periodicity, using a basis size of \verb|nbasis = 50| and without incorporating roughness penalties. The resulting smoothed functional data were subsequently used to fit the fPCA. Conversely, the fPCs themselves were expanded on a coarser basis consisting of B-splines with \verb|nbasis = 7|, thereby allowing for up to 7 fPCs before capturing the entire variance of the data. In selecting the number of fPCs to retain, we opted to include those necessary to explain at least $\delta=0.90$ of the total variance in the temperature data. Notably, this resulted in a number of fPCs equal to 5 in our experiment.

Following this analysis, we observed that our functional regression model for maize data comprises a total of $79+2+5=86$ scalar parameters. With the identification of the fPCs now complete, we proceed with the various estimation techniques presented previously. Specifically, we focus on estimating the only functional coefficient $\gamma$, reporting three distinct estimates $\hat{\gamma}^{e}(\tau)$ for each value of $\tau \in \{0.1, 0.9\}$ (when relevant) to investigate the influence of temperatures on the tails of the distribution as well.

\begin{figure}
	\centering
	\includegraphics[width=0.6\linewidth]{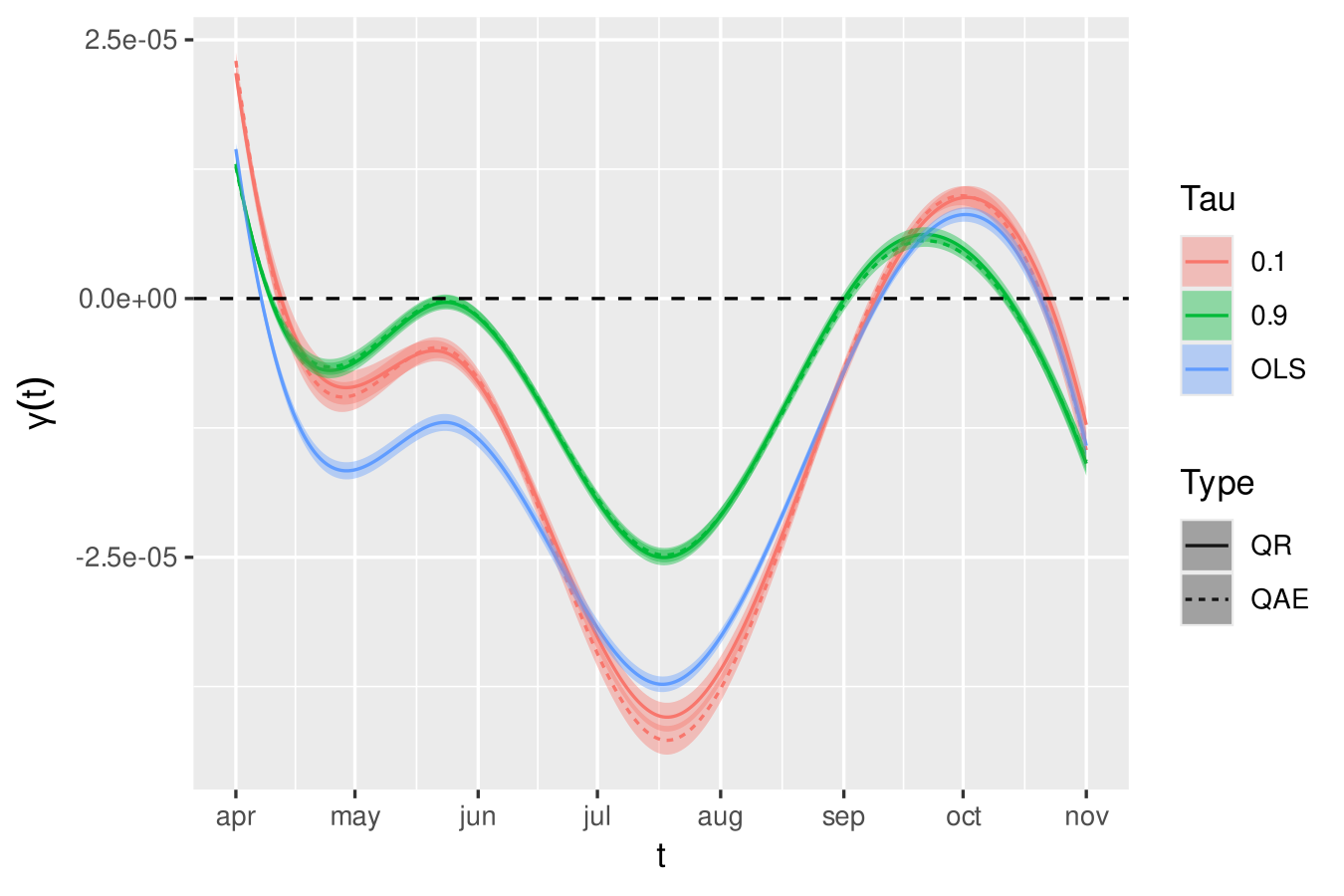}
	\caption{Estimates of the temperature functional coefficient with 95\% confidence bands. The standard OLS regression coefficient is shown in blue, while the quantile regression coefficients for $\tau = 0.1$ are shown in red and those for $\tau = 0.9$ are shown in green. As regards the quantile regression estimators alone, the estimate using simple QR is represented by a solid line, while that relating to the QA estimator is represented by a dashed line. For the QA approach the neighbourhood considered is $\mathcal{N}(\tau) = \{\tau\pm 0.025,\tau\pm 0.05\}$.}
	\label{fig:maizetemp}
\end{figure}

Figure~\ref{fig:maizetemp} presents the estimates of the functional coefficients for various estimation techniques. A primary observation is that these methods yield consistent results with respect to one another, indicating a degree of agreement among them. Notably, all coefficient functions exhibit their most negative peak during the hot season, specifically in July. Furthermore, each coefficient is characterized by negative values throughout the period from June 1st to August 31st, a period where temperatures above average have been shown to exert the most detrimental effects on maize yields. These findings align with existing literature \citep{Schlenker2009, Ortiz-Bobea2019, Ortiz-Bobea2021}. 

A second observation of particular interest is that the coefficient corresponding to the quantile 0.9 exhibits a reduced negative effect compared to the quantile 0.1, which is situated near the average value. This phenomenon can be attributed to the fact that technologically advanced farms are able to implement strategies mitigating the adverse effects of extremely high temperatures, whereas less efficient farms are disproportionately impacted by heat, as evidenced by the convergence of the red and blue curves from June onwards. These findings are in line with \citep{Chavas2019}, which analyzes seven Italian provinces in the 1900-2014 period using a quantile regression. 

Finally, it is worth noting that all coefficient functions exhibit mild positivity in April and between September and October, suggesting that temperatures above average may have a positive impact on yields during colder stages of the maize growing season \citep{Ortiz-Bobea2021}.

Furthermore, we can interpret the estimate of the functional coefficient $\gamma(t)$ as follows: suppose we observe, in a province $j$ during year $i$, a temperature curve equal to $\Delta\verb|Temp|^{t_0, t_1}_{i,j} (t)  = \verb|Temp|_{i,j}(t) + \Delta T\,\mathbf{1}_{[t_0, t_1]}(t)$. This corresponds to a temperature profile increased by $\Delta T$ degrees only during the subinterval $[t_0, t_1]$. With this assumption, we note that the difference in the response variable is equal to:
\begin{equation}
	\Delta Y_{i,j} = \Delta\log(\verb|Yield|_{i,j})= \Delta T\int_{t_0}^{t_1} \gamma(t)\, dt
\end{equation}
This gives the yield ratio of the two observations, assuming everything else is constant:
\begin{equation}
	r(\texttt{Yield}_{i,j}) = \frac{\texttt{Yield}\bigl(\Delta\texttt{Temp}^{t_0, t_1}_{i,j}\bigr)}{\texttt{Yield}(\texttt{Temp}_{i,j})}= \exp\biggl(\Delta T\int_{t_0}^{t_1} \gamma(t)\, dt\biggr)
\end{equation}

In particular, for our analysis, if we set $t_0$ to be June 1st and $t_1$ to be August 31st and $\Delta T = 1^{\circ}\text{C}$ we get the following results

\begin{table}[htp]
	\centering
	\begin{tabular}{l|cc}
		\toprule
		Estimator & $\Delta Y_{i,j}$ & $r(\texttt{Yield}_{i,j})$\\
		\midrule
		OLS	 & -0.0577 & 0.9440	\\	
		QR $\tau = 0.1$ & -0.0594 & 0.9424	\\	
		QR $\tau = 0.9$ & -0.0326 & 0.9679\\
		\bottomrule
	\end{tabular}
	\caption{The table provides the values of the difference in the maize response variable and yield ratio when passing from $\texttt{Temp}_{i,j}(t)$ to $\Delta\texttt{Temp}^{t_0, t_1}_{i,j}(t)$, with $\Delta T =1^{\circ}C$, $t_0 =$ June 1st and $t_1 =$ August 31st.}
	\label{tab:maizeeffects}
\end{table}

We conclude that, given the fitted model, one additional degree throughout June and August results in a reduction between 3.3\% and 5.8\% of maize yield. For comparison, the closest available estimate from \citep{Chavas2019} indicates that one additional degree over the entire year leads to an average yield reduction of 1.04\%.\footnote{This value has been computed using the average temperature across provinces and the estimated coefficients reported in Tables 1 and 3 of \citep{Chavas2019}.}

Finally, given that our analysis reveals minimal discrepancy between the QR and QA estimates, we deemed it unnecessary to test the hypothesis required for sound use of the QA estimate. In this case, we can restrict our attention solely to the QR estimate, which is sufficient for the two chosen quantiles. 

It is noteworthy that not only does our model provide a meaningful interpretation of the temperature functional coefficient and the effects of intra-season temperatures on the final crop yield, but it also exhibits strong predictive capabilities. Specifically, the OLS estimate yields an adjusted $R^2$ value of 0.9946 with a correlation between true and predicted responses of 0.901.


\subsection{Wheat yields analysis}

We proceed with our analysis by focusing on the wheat data, which presents distinct characteristics compared to the maize dataset. As wheat is a non-irrigated crop, we decided to incorporate precipitation data into our analysis. Notably, the precipitation series is characterized by numerous zeros, making it challenging to incorporate into an FDA framework, even after centering. 

Furthermore, it is well-documented that precipitation tends to have a less immediate impact on crops compared to temperature fluctuations. In light of this, we employ a transformation of the precipitation series to capture its effects more effectively. To achieve this, we apply a convolution with a linear filter $w$, with the aim of obtaining a moving average of the precipitation series.

\[
\verb|Prec|_{i,j}^w(t) = \verb|Prec|_{i,j}(t) * w(t)
\]
The window function $w(t)$ is defined in the following way
\[
w(t) = \begin{cases}
	1 & 0 \le t \le W\\
	0 & t < 0\text{ or }t > W
\end{cases}
\]

Following the convolution step, we obtain a transformed precipitation series $\verb|Prec|_{i,j}^w(t)$, which can be expressed as the cumulative sum over a window of size $W$ extending into the past: $\verb|Prec|_{i,j}^w(t) = \sum_{s = t - W}^t\verb|Prec|_{i,j}(s)$. 

Notably, if the window size $W$ is allowed to increase up to the length of the series, the transformed series reduces to the standard cumulative sum. In our experiments, we explore a range of values for W, and ultimately find that the transformation yielding the most interpretable coefficients is indeed the cumulative sum of the precipitations. 

With this preprocessing step complete, we proceed with the analysis of the wheat model for yields: 

\begin{equation}
	\label{eqn:modelwheat}
	Y_{i,j} = \alpha_j + \beta_1 (i - i_0)+ \beta_2 (i - i_0)^2 + \int_{0}^T \gamma^1(t) (\verb|Temp|_{i,j} - \verb|Temp|_{\bullet,\bullet})(t) dt + \int_{0}^T \gamma^2(t) (\verb|Prec|^w_{i,j} - \verb|Prec|^w_{\bullet,\bullet})(t) dt + \varepsilon_{i,j}
\end{equation}

In a manner analogous to our previous analysis, we consider a dataset encompassing $N=72$ years (1952-2023) and $M=68$ provinces for which a complete series of yields is available. Consistent with our prior approach, temperature data were expanded onto a Fourier basis with \verb|nbasis = 50| functions. For the transformed precipitation series, given its increasing nature, we employ a polygonal basis to achieve a linear interpolation of the sample data. 

Subsequently, we apply functional principal component analysis (fPCA) using a B-spline basis with \verb|nbasis = 6| to model the functional principal components (fPCs) for both covariates. Our results indicate that four fPCs for temperatures and two fPCs for precipitations are necessary to capture at least a fraction $\delta=0.9$ of the total variance. Consequently, the wheat model incorporates a total of 68+2+4+2=76 parameters. 

We obtain estimates for the two functional coefficients, $\gamma^{1, e}(t, \tau), \gamma^{2, e}(t, \tau)$, using each estimation approach $e$, considering relevant quantile values $\tau \in \{0.1, 0.9\}$ when applicable. 

\begin{figure}
	\centering
	\begin{subfigure}{0.49\textwidth}
		\includegraphics[width=\textwidth]{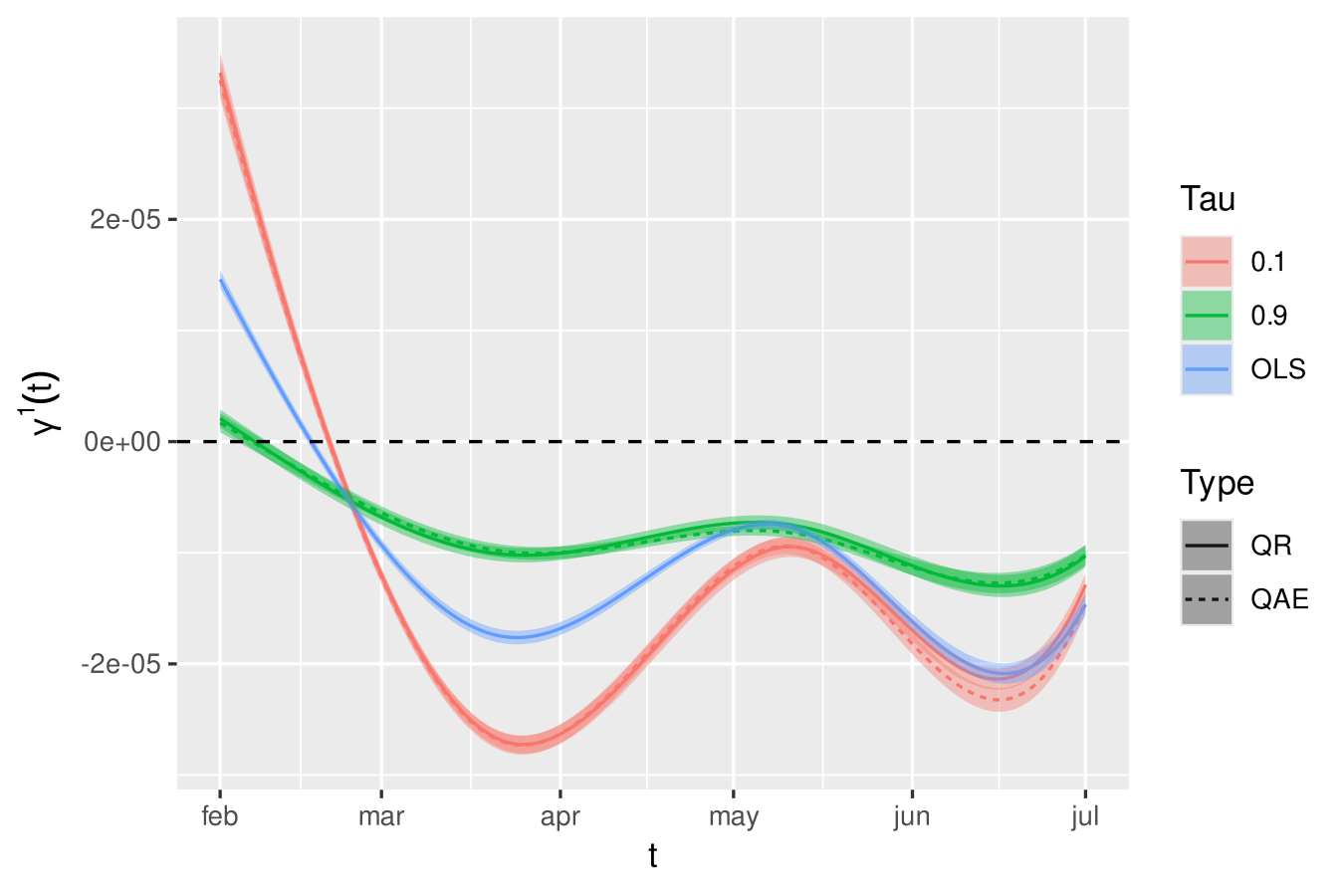}
		\caption{}
		\label{fig:wheattemp}
	\end{subfigure}
	\hfill
	\begin{subfigure}{0.49\textwidth}
		\includegraphics[width=\textwidth]{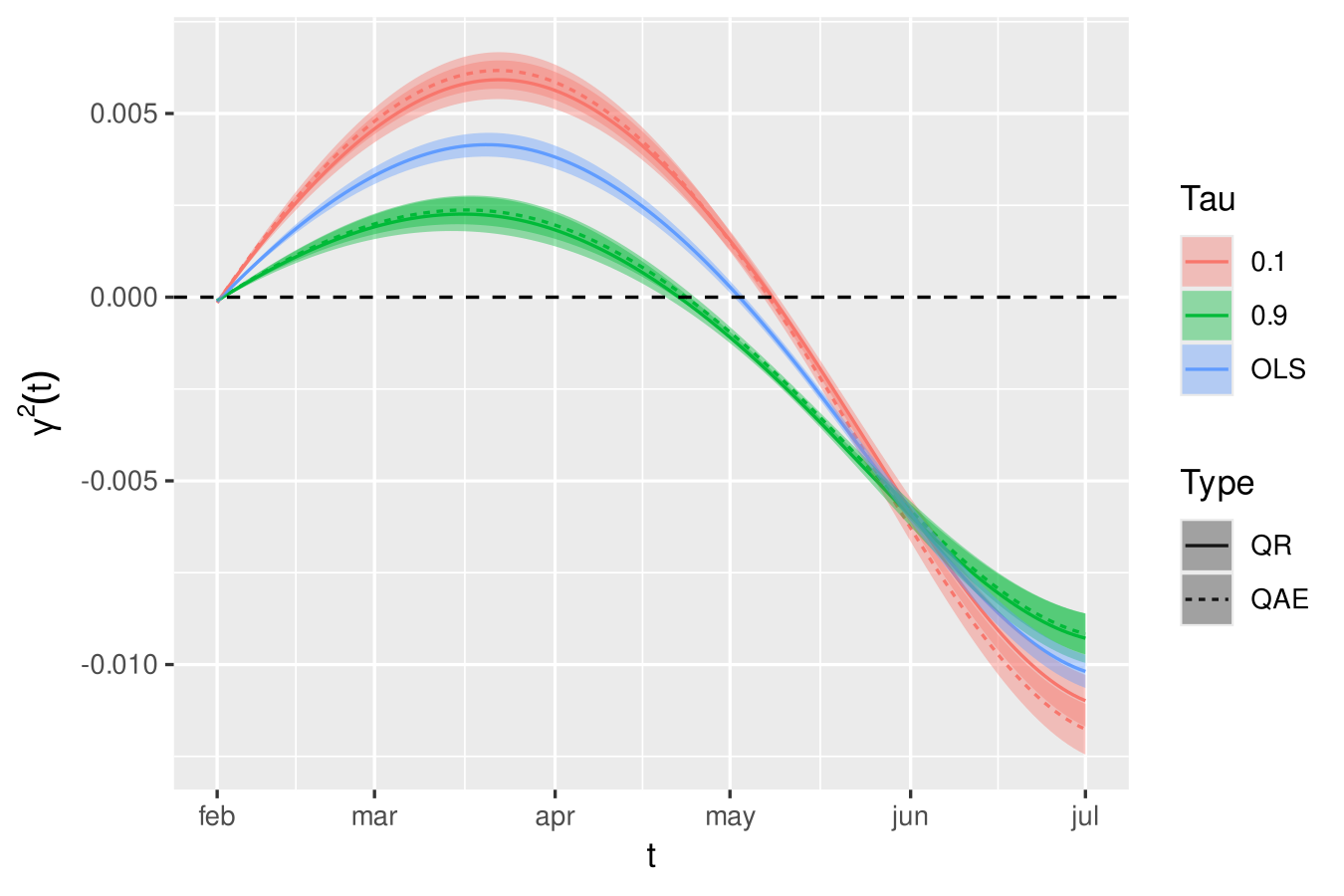}
		\caption{}
		\label{fig:wheatprec}
	\end{subfigure}
	\caption{Estimates of the functional coefficients of temperature (panel (a)) and precipitation (panel (b)) with 95\% confidence bands. The standard OLS regression coefficient is shown in blue, while the quantile regression coefficients for $\tau = 0.1$ are shown in red and those for $\tau = 0.9$ are shown in green. As regards the quantile regression estimators alone, the estimate using simple QR is represented by a solid line, while that relating to the QA estimator is represented by a dashed line. For the QA approach the neighbourhood considered is $\mathcal{N}(\tau) = \{\tau\pm 0.025,\tau\pm 0.05\}$.}
	\label{fig:wheatcoeff}
\end{figure}

Figure~\ref{fig:wheatcoeff} presents the estimated functional coefficients of the climate covariates for wheat. Our analysis reveals that high temperatures have a negative impact on wheat production, particularly during the period spanning from late March to early April, which corresponds to the critical flowering phase of the crop. This suggests that temperatures above the mean within this time window may compromise growth and ultimately lead to reduced yields. 

Furthermore, we observe a consistent ordering of the coefficients: the curve corresponding to the 0.9 quantile consistently remains above the others, while the one associated with the 0.1 quantile remains at the bottom. This finding supports the notion that more productive farms are less susceptible to the negative impacts of abrupt temperature changes compared to less productive farms. 

Similarly to what we did in the previous section, we can quantify the effects of variation in temperatures on the response variable. Regarding temperatures, and fixing $t_0 =$ March 1st, $t_1 =$ May 31st and $\Delta T = 1^\circ C$ we obtain the results reported in Table \ref{tab:wheateffects}.

\begin{table}[htp]
	\centering
	\begin{tabular}{l|cc}
		\toprule
		Estimator & $\Delta Y_{i,j}$ & $r(\texttt{Yield}_{i,j})$\\
		\midrule
		OLS	 & -0.0203 & 0.9799	 \\	
		QR $\tau = 0.1$ & -0.0310 & 0.9695	\\	
		QR $\tau = 0.9$ & -0.0131 & 0.9870	\\
		\bottomrule
	\end{tabular}
	\caption{The table provides the values of the difference in the wheat response variable and yield ratio when passing from $\texttt{Temp}_{i,j}(t)$ to $\Delta\texttt{Temp}^{t_0, t_1}_{i,j}(t)$ (assuming constant precipitations), with $\Delta T =1^{\circ}C$, $t_0 =$ March 1st and $t_1 =$ May 31st.}
	\label{tab:wheateffects}
\end{table}
Once more, we conclude that, given the fitted model and assuming no change in the precipitation variable, one additional degree throughout March and May results in a reduction between 1.3\% and 3\% of wheat yield. Again, for comparison, estimates from \citep{Chavas2019} indicate that one additional degree over the entire year leads to an average wheat yield reduction of only 0.03\%, once more highlighting the importance of considering seasonal effects.

Regarding precipitation, we find that the positive effect of a cumulative amount of rain above the average is confined to the period up to the end of May and particularly during the flowering phase of wheat. In this case, the hierarchy of coefficients is reversed: more productive farms are less positively impacted by precipitation above the mean, while less productive farms appear to benefit more from such a phenomenon. This can be due to the better water management practices of more efficient farms, which depend less on rainwater to achieve optimal yields. We also observe a negative effect of cumulative rains in the latter part of the growing season, particularly during the grain-filling stage, when excessive moisture can hinder the maturation of wheat, leading to lower grain quality and yield. 

To quantify the impact of precipitation variations, we propose a modified precipitation curve for province $j$ during year $i$, defined as follows: $\Delta\verb|Prec|^{w,t_0, t_1}_{i,j} (t)  = \verb|Prec|^w_{i,j}(t) + \Delta P \Bigl[\frac{t-t_0}{t_1-t_0} \Bigl(1-\frac{1}{t_1-t_0}\Bigr)+\frac{1}{t_1-t_0}\Bigr]\,\mathbf{1}_{[t_0, t_1]}(t)$. This expression introduces an additional linear term, going from $\frac{\Delta P}{t_1-t_0}$ at time $t_0$ to $\Delta P$ at time $t_1$. Since the cumulative sum of precipitation is represented by $\verb|Prec|^w_{i,j}(t)$, this modification effectively models a variation in precipitation of size $\Delta P$, evenly distributed throughout the interval $[t_0,t_1]$. The sign of $\Delta P$ determines whether the increase or decrease in precipitation is modeled.

Similarly to what has been done for temperatures, we can compute the difference in the response variable:
\begin{equation}
	\Delta Y_{i,j} = \Delta\log(\verb|Yield|_{i,j})= \Delta P\int_{t_0}^{t_1} \Bigl[\frac{t-t_0}{t_1-t_0}\Bigl(1-\frac{1}{t_1-t_0}\Bigr)+\frac{1}{t_1-t_0}\Bigr]\gamma^2(t)\, dt
\end{equation}
and, assuming everything else stays constant (including the temperature curve), we can derive the yield ratio as
\begin{equation}
	r(\texttt{Yield}_{i,j}) = \frac{\texttt{Yield}\bigl(\Delta\texttt{Prec}^{w,t_0, t_1}_{i,j}\bigr)}{\texttt{Yield}(\texttt{Prec}^w_{i,j})}= \exp\biggl(\Delta P\int_{t_0}^{t_1} \Bigl[\frac{t-t_0}{t_1-t_0}\Bigl(1-\frac{1}{t_1-t_0}\Bigr)+\frac{1}{t_1-t_0}\Bigr]\gamma^2(t)\, dt\biggr)
\end{equation}

In particular, if we set $t_0 =$ February 1st, $t_1 =$ April 30th and $\Delta P = - 100 \,\text{mm}$ we obtain the results reported in Table~\ref{tab:preceffects}.

\begin{table}[h]
	\centering
	\begin{tabular}{l|cc}
		\toprule
		Estimator & $\Delta Y_{i,j}$ & $r(\texttt{Yield}_{i,j})$\\
		\midrule
		OLS &  -0.012403267	& 0.987673		\\	
		$\tau = 0.1$ &  -0.019360238	& 0.980826		\\	
		$\tau = 0.9$ &  -0.004818595	& 0.995193 \\
		\bottomrule
	\end{tabular}
	\caption{The table provides the values of the difference in the wheat response variable and yield ratio when passing from $\texttt{Prec}^w_{i,j}(t)$ to $\Delta\texttt{Prec}^{w,t_0, t_1}_{i,j}(t)$ (assuming constant temperatures), with $\Delta P =-100\,\text{mm}$, $t_0 =$ February 1st and $t_1 =$ April 30th.}
	\label{tab:preceffects}
\end{table}

A reduction of 100 mm in precipitation, which is evenly distributed between February 1st and April 30th, has a detrimental effect on yields, resulting in a yield decrease ranging from 0.5\% to 1.9\%. This specific period was chosen for analysis as it corresponds to the critical wheat growth phase, during which artificial irrigation can be effectively employed to mitigate adverse weather conditions. Furthermore, our results indicate that the magnitude of yield reduction associated with decreased precipitation is less pronounced than that associated with increased temperatures.\footnote{This result is confirmed by considering a deviation of one standard deviation for both variables.}

As in the case of the maize coefficients, the difference between QR and the QA estimates is minimal, and thus, we do not check the hypothesis to employ the QA estimate but consider only the simple QR estimate for the two quantiles chosen.

Once more, our results indicate that the wheat model of Equation~\ref{eqn:modelwheat} not only facilitates the interpretation of functional coefficients in terms of their effects during the growing season but also exhibits strong predictive capabilities. Specifically, the OLS estimate of the model yields an adjusted $R^2$ value of 0.9975, while the correlation between the true yields and the predicted yields is 0.933. 

\section{Conclusions}\label{sec5}

In this work, we presented a scalar-on-function linear regression model for predicting crop yields using both temporal and spatial scalar variables and functional weather covariates. The models were fitted using fPCA, and the functional coefficients could be obtained by considering different estimation approaches such as OLS and quantile regression. For both models considered, for maize and soft wheat respectively, the functional coefficients of temperature allow us to infer a negative impact of anomalous temperatures occurring at crucial stages of the crop growing seasons. The models, in addition to offering interesting interpretations of the effects of weather covariates, are also quite predictive. 

For maize, we considered a single weather variable, namely temperature, due to the irrigated nature of this crop; for wheat, on the other hand, we considered two variables, namely temperature and precipitation. Our analysis shows that high temperatures depress maize yields during the core summer months (June–August), while they tend to be beneficial at the beginning and end of the growing season (April and October). For soft wheat, by contrast, temperature spikes exert a predominantly negative influence, with the most pronounced effects occurring from late March to early April. Precipitation also plays a season-dependent role: increased rainfall enhances wheat yields in the early stages of growth but becomes detrimental later in the season. 
These findings underscore the importance of accounting for the timing of weather anomalies and provide valuable information for adaptation to climate change. In particular, they suggest that more targeted adjustments of agronomic inputs, such as irrigation, could help buffer crops against weather-related stress precisely when such interventions are most effective.

A natural extension of this work concerns the joint effects of weather variables. In this study, we examined temperature and precipitation individually, but future research should explore their interactions, particularly given that the impact of abnormal temperatures may be amplified under low-rainfall conditions. Another promising direction involves addressing boundary effects arising from the use of B-splines in the basis expansion of the regression coefficients. Although we did not emphasize this aspect, our primary interest being the central portion of the growing season, it may be worthwhile to consider methods that mitigate such edge distortions. In particular, the application of group-Lasso penalties~\citep{bernardi2023} or related regularization strategies could help reduce the influence of boundary artifacts and improve model stability.

\section*{Acknowledgments}
AM and AO acknowledge partial financial support by Fondazione Cariplo under the project \emph{Data Science Approach for Carbon Farming Scenarios (DaSACaF)}. GB and AM acknowledge partial support to INDAM (Istituto Nazionale di Alta Matematica), Italy. Computational resources have been provided by the core facility INDACO, Università degli Studi di Milano. 

\subsection*{Author contributions}

G.B., A.M. developed the statistical modelling; A.O., P.N. provided the data and conceived the agro-environmental problem; G.B. wrote the codes and ran the experiments; G.B., P.N. processed the data; A.M., A.O. supervised the experiments and acted as coordinators. All authors wrote and reviewed the manuscript.

\bibliographystyle{apalike}
\bibliography{references.bib}%

\end{document}